# Search for Intermediate Mass Magnetic Monopoles and Nuclearites with the SLIM experiment

S. Cecchini[1,7], T. Chiarusi[1], D. Di Ferdinando[1], M. Cozzi[1], M. Frutti[1], G. Giacomelli[1], A. Kumar[1,8], S. Manzoor[1,4], J. McDonald[3], E. Medinaceli[1,6], J. Nogales[6], L. Patrizii[1*], J. Pinfold[3], V. Popa[1,5], I.E. Qureshi[4], O. Saavedra[2], G. Sher[4], M.I. Shahzad[4], M. Spurio[1], R. Ticona[6], V. Togo[1], and A. Velarde[6]

[1] Dip. Fisica dell'Universita' di Bologna and INFN, 40127 Bologna, Italy
[2] Dip. Fisica Sperimentale e Generale dell'Universita' di Torino and INFN, 10125 Torino, Italy
[3] Centre for Subatomic Research, Univ. of Alberta, Edmonton, Alberta T6G 2N4, Canada
[4] PRD, PINSTECH, P.O. Nilore, Islamabad, Pakistan
[5] Institute for Space Sciences, 77125 Bucharest, Romania
[6] Laboratorio de Fisica Cosmica de Chacaltaya, UMSA, La Paz, Bolivia
[7] INAF/IASF Sez. di Bologna, 40129 Bologna, Italy
[8] Dept. of Physics, SHSL-Central Institute of Eng. & Tech., Longowal-148 106 India

**Abstract**
SLIM is a large area experiment (440 m$^2$) installed at the Chacaltaya cosmic ray laboratory since 2001, and about 100 m$^2$ at Koksil, Himalaya, since 2003. It is devoted to the search for intermediate mass magnetic monopoles ($10^7$-$10^{13}$ GeV/c$^2$) and nuclearites in the cosmic radiation using stacks of CR39 and Makrofol nuclear track detectors. In four years of operation it will reach a sensitivity to a flux of about $10^{-15}$ cm$^{-2}$ s$^{-1}$ sr$^{-1}$. We present the results of the calibration of CR39 and Makrofol and the analysis of a first sample of the exposed detector.

___________________
[*]Corresponding author. Tel.: +39-051-2095247; fax: +39-051-2095269
E-mail address: patrizii@bo.infn.it

## 1. Introduction

Grand Unified Theories (GUT) of electroweak and strong interactions predict the existence of superheavy Magnetic Monopoles (MMs) with masses larger than $10^{16}$ GeV. They would have been produced at the end of the GUT epoch, at a mass scale ~$10^{14}$ GeV and the cosmic time $10^{-34}$ s (Preskill, J., 1984; Giacomelli, G., 1984; Groom, D.E.N., 1986). The MACRO experiment has set the best limits on GUT MMs for $4 \cdot 10^{-5} < \beta = v/c < 0.5$ (Ambrosio, M. et al., 2002).

Intermediate Mass Magnetic Monopoles (IMMs) may have been produced in later phase transitions in the Early Universe (Kephart T.W. and Shafi Q. 2001). Relativistic

MMs with intermediate masses, $10^7 < m_M < 10^{13}$ GeV, could be present in the cosmic radiation. IMMs could be accelerated to large values of $\gamma$ in one coherent domain of the galactic magnetic field. It has been speculated that very energetic IMMs could yield the highest energy cosmic rays (Kephart T.W. and Weiler T.J., 1996, Escobar C.O. and Vasquez R.A. 1999). Thus one would have to look for $\beta \geq 0.1$ fast, heavily ionizing MMs.

Detectors underground, underwater and under ice would mainly have sensitivity for IMMs coming from above (Derkaoui, J. et al., 1998). Detectors at the Earth surface could detect IMMs coming from above if they have masses larger than $10^5$-$10^6$ GeV (Cecchini S. et al, 2001a); lower mass IMMs may be searched for with detectors located at high mountain altitudes or in balloons and satellites. Fig. 1a shows the experimentally accessible region in the search for IMM's: the minimal velocity at the entry point in the atmosphere versus the monopole mass, for detectors located at different altitudes.

Nuclearites, (strange quark matter, SQM) should consist of aggregates of *u, d* and *s* quarks in approximately equal proportions (Witten A., 1986; De Rujula A. and Glashow S.L., 1984). The SQM is a color singlet, thus it may have only integer electric charges. The finite SQM should have a positive charge, and the overall neutrality of nuclearites is ensured by an electron cloud, forming a sort of atom. SQM has been proposed as the ground state of QCD. Nuclearites should be stable for all baryon numbers in the range between ordinary heavy nuclei and neutron stars (A $\sim 10^{57}$). Fig. 1b shows the experimentally accessible region in the search for nuclearites. Much lower mass nuclearites (A $\sim$ few $10^2$) could also reach the SLIM altitude coming from above, if one assumes some particular interaction models with the nuclei in the atmosphere (Wilk, G. and Wlodarczyk, Z., 1996; Banerjee, S. et al., 2000).

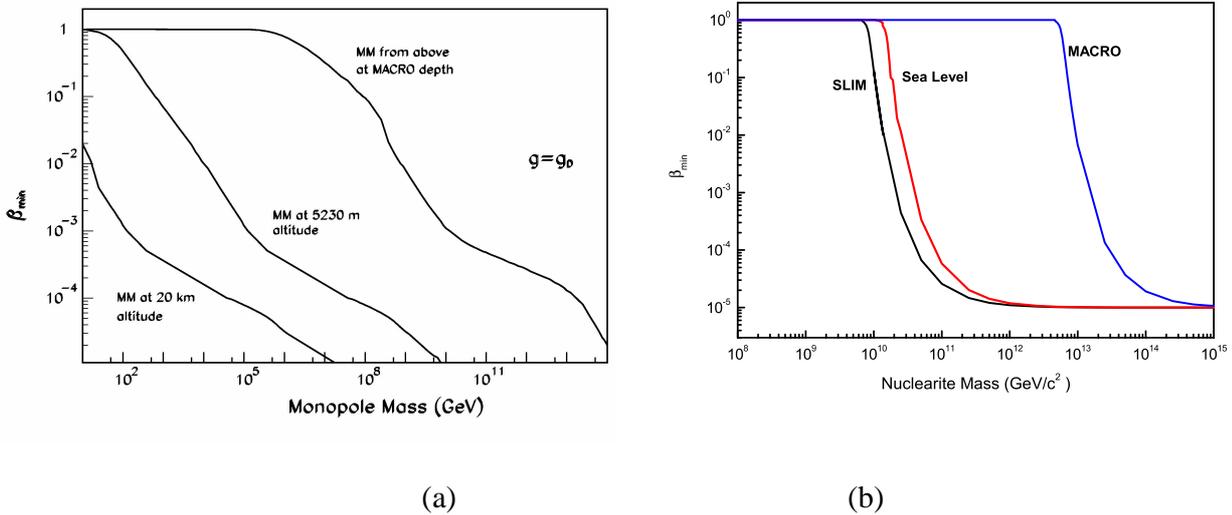

(a)                              (b)

Fig. 1. Accessible regions in the plane (mass, β) for (a) MMs and (b) nuclearites coming from above for experiments at high altitudes and underground.

Q-balls are super-symmetric coherent states of squarks, sleptons and Higgs fields,

predicted by minimal super-symmetric generalizations of the Standard model (Coleman, S., 1985; Kusenko, A. and Shaposhnikov, A., 1998) they could have been produced in the early universe; their interaction with matter is expected to be, at least qualitatively, similar to that of nuclearites.

MMs, nuclearites and Q-balls could be components of the galactic cold dark matter.

In this paper we present the SLIM (Search for LIght magnetic Monopoles) experiment that was designed to search for fast IMMs, nuclearites and Q-balls with nuclear track detectors at the Chacaltaya high altitude laboratory (5230 m above sea level) (Bakari, D et al., 2000; Cecchini, S. et al., 2001a) . The high altitude exposure will allow detection of the above mentioned particles even if they have strong interaction cross sections which could prevent them from reaching the earth surface (Rybczynski, M. et al., 2001). We also discuss the new etching procedures of the SLIM nuclear track detectors (NTDs) and report the preliminary flux limit obtained by analyzing the first 60 $m^2$ of the detector.

## 2. Experimental

The SLIM apparatus at Chacaltaya consists of 440 $m^2$ of CR39 and Makrofol nuclear track detectors. The installation began in March 2000 and was completed in July 2001. Further 100 $m^2$ were installed at Koksil (Himalaya) since 2003.

The detector is organized in modules of 24 cm x 24 cm, each made of 3 layers of CR39 (1.4 mm thick), 3 layers of polycarbonate (Makrofol, 0.5 mm thick) and of an aluminium absorber (1 mm thick); each module is sealed in an aluminized plastic bag filled with dry air. Since the atmospheric pressure at Chacaltaya is ~ 0.5 atm, we made a test in which some envelopes filled with 1 atm of air were sealed and placed in a chamber at a pressure of about 0.3 atm for three weeks; no leakage was detected in any of them.

From our experience with the MACRO experiment, where the same CR39 material was used, we know that such material does not suffer from "aging effects" for exposure times as long as 10 years, that is, there is no appreciable dependence of the detector response on the time elapsed between the date of production and the passage of the particle (Cecchini, S. et al., 2001b).

The planned exposure time of the SLIM detector is 4 years.

## 3. Background estimates. Etching procedures

We performed tests by exposing nuclear track detectors in Bologna and at the Chacaltaya mountain station, in order to study the effects of possible backgrounds and of possible climatic conditions.

Preliminary results of radon concentration in the experimental rooms at Chacaltaya were also obtained by using E-PERM radon dosimeters. The radon activity in different locations around SLIM, was found to be about 40 - 50 $Bq/m^3$. From our experience with the MACRO experiment at LNGS, we conclude that such levels of radon activity are not a problem for the experiment.

We made also preliminary measurements of the flux of cosmic ray neutrons with energy 1 < $E_n$ < 20 MeV, in the vicinity of the SLIM detector, using bubble counters and $BF_3$ gaseous detectors. We obtained $\Phi_n = (1.7 \pm 0.8)\ 10^{-2}\ cm^{-2}\ s^{-1}$, in agreement with other reported neutron flux data at such altitudes (Grieder, P.K.F., 2001). These neutrons by interacting inside the detector could cause background tracks on the sheets and affect the time needed for the scanning and analysis, and the scan efficiency. For these reasons we have studied new improved etching procedures, based on the addition of ethyl alcohol to water solutions of NaOH and KOH. These tests have shown an increase of the signal/background ratio with a drastic reduction the background tracks on the detector sheets. At the end of these tests we have defined new "strong" and "soft" etching conditions (Manzoor, S. et al., 2004):

- strong etching conditions - 8N KOH + 1.25% Ethyl alcohol at 77 °C for 30 hours. The strong etching is used for the first CR39 sheet in each module, in order to produce large tracks, easier to detect during scanning.
- soft etching conditions - 6N NaOH + 1% Ethyl alcohol at 70 °C for 40 hours. The soft etching is applied to the other CR39 layers in a module, if a candidate track was found in the first layer. It allows reliable measurements of the restricted energy loss and direction of the incident particle.

**4. Calibrations**

The CR39 and Makrofol detectors were calibrated with 30 A GeV $Pb^{82+}$, and 160 A GeV $In^{49+}$ beams at the CERN SPS, and 1 A GeV $Fe^{26+}$ at the BNL AGS. The layout of the exposures is sketched in Fig. 2. The sheets behind the target can detect both primary ions and their nuclear fragments with decreasing charge from $Z = Z_{beam}$ to $Z_{threshold}$. The base areas of the "post-etch cones" were measured with an automatic image analyzer system (Noll, A. et al., 1988).

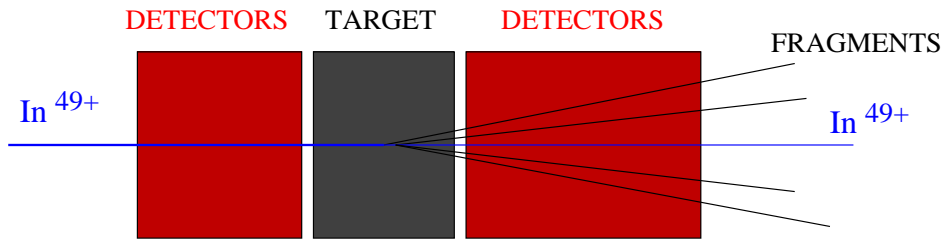

Fig. 2. Scheme of the exposure to $In^{49+}$ 160 A GeV ions of the calibration stacks. Both CR39 and Makrofol NTDs were used.

The base area distribution of the $In^{49+}$ ions and of and their fragments in CR39 etched with "soft" etching condition is shown in Fig. 3. Fig. 4 shows the calibrations of our detectors plotting the experimental values of p-1 ($p = v_T/v_B$, where $v_T$ and $v_B$ are the track and bulk etching velocity, respectively) vs. the Restricted Energy Loss (REL) for "soft" and "strong" etching.

For "soft" etching the threshold in CR39 is $Z/\beta \sim 7$ which corresponds to REL $\sim 50$ MeV cm$^2$ g$^{-1}$. For strong etching the threshold is at $Z/\beta \sim 14$ which corresponds to REL $\sim 200$ MeV cm$^2$ g$^{-1}$.

The Makrofol polycarbonate has a higher threshold ($Z/\beta \sim 50$). We discuss the calibration of this detector in an accompanying paper (Manzoor, S. et al., 2004).

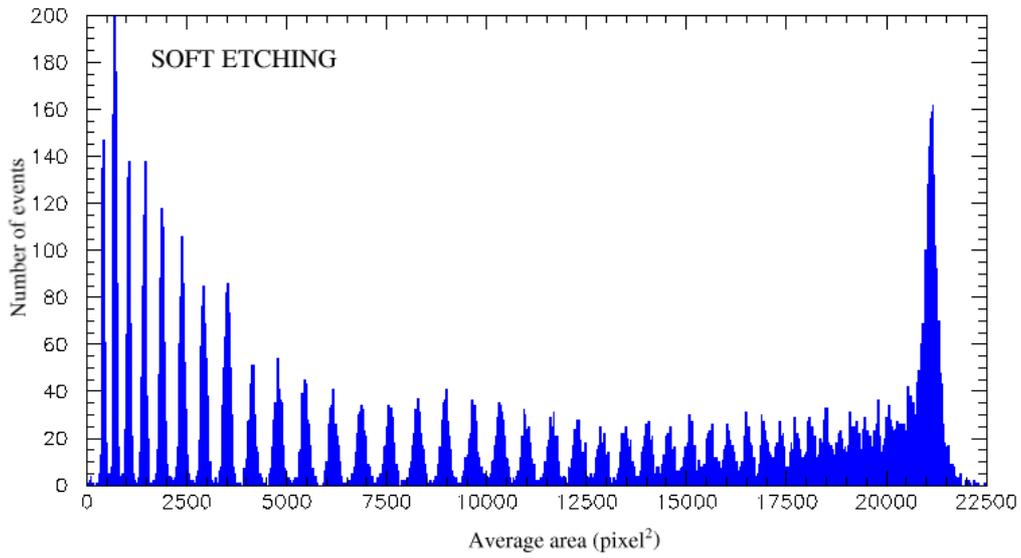

Fig. 3. Distribution of the base areas of the etched cones produced in CR39 by relativistic In$^{49+}$ ions and their fragments. The base areas are averaged over two faces. The CR39 was etched in "soft" conditions.

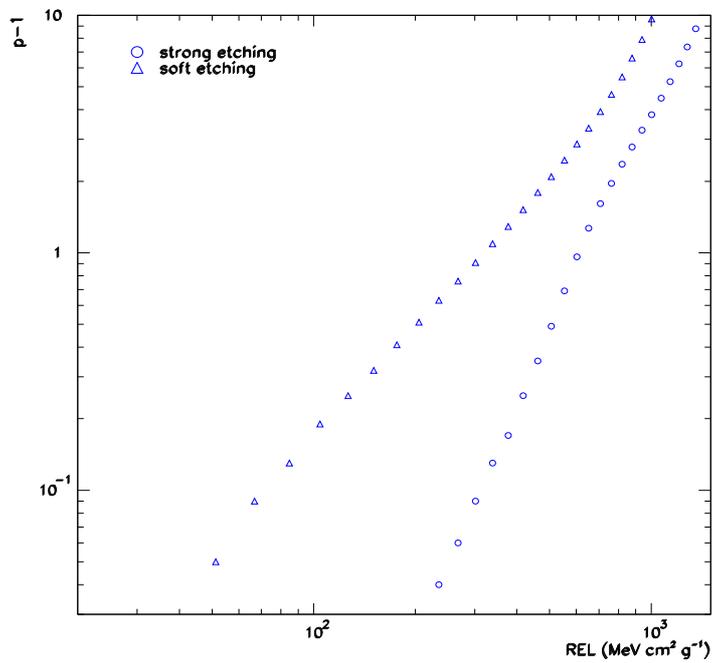

Fig. 4. Reduced track etch rate (p-1) vs REL for the CR39 detectors (exposed to the indium ion beam) etched in the "soft" (triangles) and "strong" (circles) etching conditions.

With the aforementioned etching conditions the CR39 allows the detection of MMs with one unit Dirac charge ($g=g_D$), for $\beta$ around $10^{-4}$ and for $\beta >10^{-3}$, the whole $\beta$-range of $4 \times 10^{-5} < \beta < 1$ for MMs with $g \geq 2\ g_D$ and for dyons.

The Makrofol is useful for the detection of fast MMs. Nuclearites with $\beta \sim 10^{-3}$ can be detected by both CR39 and Makrofol.

If nuclearites of relatively low mass could be accelerated as ordinary comic ray nuclei, they could achieve higher velocities and still reach the detector level.

## 5. Analysis

When the exposure at Chacaltaya is completed, the modules are brought back to Italy and analyzed in the Bologna laboratory. Three fiducial holes of 2 mm diameter are drilled in each module with a precision drilling machine (the position of the holes is defined to within a 100 µm accuracy). The bags are open and the detectors are labelled and their thickness is measured in 9 uniformly distributed points

The analysis of a SLIM module begins by etching the uppermost CR39 sheet using "strong" conditions in order to quickly reduce its thickness from 1.4 mm to ~0.4 mm. Since MMs, nuclearites and Q-balls have a constant REL through the stack, the signal we are looking for is a hole or a biconical track with the two base-cone areas equal within the experimental uncertainties. After the strong etching the sheets are scanned twice by different operators with a 3× magnification optical lens looking for any possible optical inhomogeneity. An optical inhomogeneity is usually a particle track or a defect of the material. The double scan guarantees an efficiency of ~ 100% for finding a possible signal.

The detected inhomogeneities are then observed with a stereo 20-40 × microscope and classified either as defects or particle tracks requiring further analyses. Each candidate track is further observed under an optical microscope with 3.2×25 or 6.3×25 magnification and the minor *axes* of the base-cone ellipse in the front ($A_{front}$) and back ($A_{back}$) sides are measured. We require that $|A_{front} / A_{back} - 1| \leq 3$ times the standard deviation of the difference.

In order to compute the p-values and incident angles $\theta$ for the front and back sides, the track major axes are also measured. Finally a track is defined as a "candidate" if p and $\theta$ on the front and back sides are equal to within 15%. For each candidate the azimuth angle $\varphi$ and its position P referred to the fiducial marks are determined. The uncertainties $\Delta\theta$, $\Delta\varphi$ and $\Delta P$ define a "coincidence" area (<0.5 cm$^2$) around the candidate expected position in the other layers (Fig. 5).

The lowermost CR39 layer is then etched in the "soft" conditions, and an accurate scan under an optical microscope with high magnification. is performed in a square region around the candidate expected position, which includes the "confidence" area. If a two-fold coincidence would be detected, also the CR39 middle layer would be analyzed as the lowermost one.

Up to now no two-fold coincidence has been found, that is no magnetic monopole, nuclearite or Q-ball candidate was detected.

## 6. Conclusions

A small quantity of exposed SLIM modules was removed and processed (mainly for testing the procedures). The analyzed area is ~ 60 m$^2$, with an average exposure time of 2.4 years. No candidate survived the tests, so the 90% C.L. flux upper limit for fast IMM's and nuclearites coming from above is at the level of $2 \cdot 10^{-14}$ cm$^{-2}$ sr$^{-1}$ s$^{-1}$ (see Fig. 6).

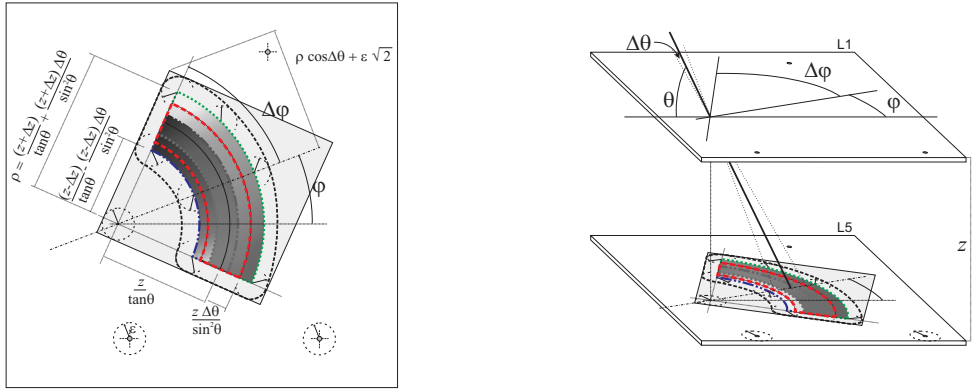

Fig. 5. Illustration of the procedure used in order to define the "confidence" area in which the possible continuation inside a SLIM module of a candidate track is searched for.

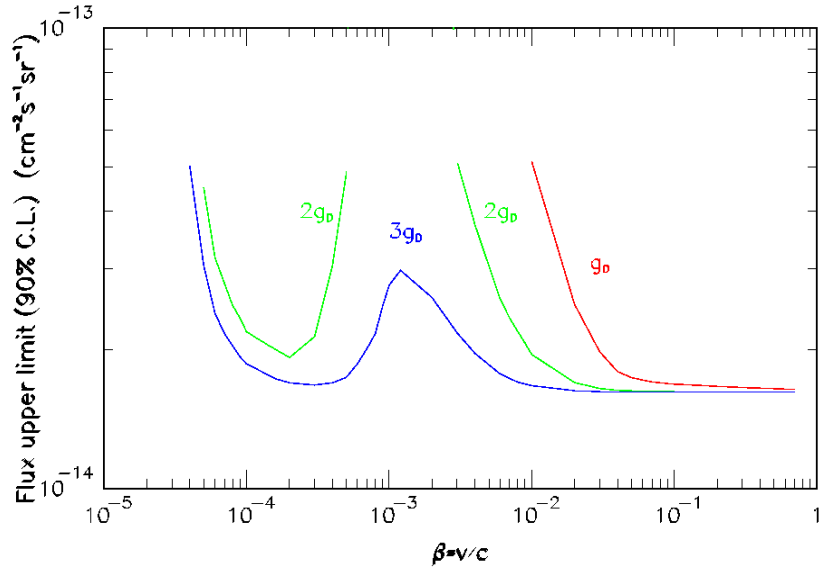

Fig. 6. The 90% C.L. upper limits to the flux of IMMs with $g=g_D$, $2\,g_D$ and $3g_D$ magnetic charges, obtained by analyzing 59 m$^2$ of the SLIM CR39 detectors exposed for 2.4 years.

The analysis of the full detector has now started. In four years of operation SLIM will reach a sensitivity of $10^{-15}$ cm$^{-2}$ sr$^{-1}$ s$^{-1}$ for $\beta \geq 10^{-2}$ and IMMs with $10^7 < m_{IMM} < 10^{13}$ GeV, for nuclearites and charged Q-balls with galactic velocities.

**Acknowledgements**

We acknowledge the collaboration of E. Bottazzi, L. Degli Esposti, R. Giacomelli and C. Valieri of INFN, Sez. Bologna, and the Chacaltaya High Altitude Laboratory technical staff.